\documentclass[]{svjour3}

\usepackage[numbers]{natbib}
\makeatletter
\DeclareMathSizes{\@xiipt}{\@xiipt}{7.5}{7}
\makeatother
\usepackage{type1cm}
\usepackage[usenames,dvipsnames]{color}
\usepackage{dsfont}
\makeatletter
\def\eqalign#1{\null\,\vcenter{\openup\jot\m@th
  \ialign{\strut\hfil$\displaystyle{##}$&$\displaystyle{{}##}$\hfil
      \crcr#1\crcr}}\,}
\makeatother

\makeatletter
\renewcommand{\paragraph}{%
\@startsection{paragraph}{4}%
{\z@}{1.25ex \@plus 1ex \@minus .2ex}{-1em}%
{\reset@font \normalsize \itshape}%
}
\makeatother
\usepackage{amsmath,amssymb}
\usepackage{amsopn}
\DeclareMathOperator{\arsinh}{arsinh}
\usepackage{graphicx}

\newcommand{\secref}[1]{section~\ref{#1}}
\newcommand{\figref}[1]{figure~\ref{#1}}

\usepackage[varg]{txfonts}

\newcommand{\added}[1]{\textcolor{red}{#1}}

\renewcommand{\added}[1]{\textcolor{black}{#1}}

\begin{document}

\title{An
analytic solution to the equations of the motion of a point mass with quadratic resistance
and generalizations}

\titlerunning{Motion of point mass with quadratic resistance and generalizations}

\author{Shouryya Ray \and Jochen Fr\"{o}hlich}
\institute{Shouryya Ray \and Jochen Fr\"{o}hlich \at Institut f\"{u}r Str\"{o}mungsmechanik, Technische
Universit\"{a}t Dresden, \\ George-B\"{a}hr-Stra{\ss}e 3c, D-01062 Dresden, Germany\\ \email{Shouryya.Ray@mailbox.tu-dresden.de}}

\maketitle

\begin{abstract}
The paper \added{is devoted to} the motion of a body in a fluid under the influence of
gravity and drag. Depending on the regime considered, the drag force can
exhibit a linear, quadratic or even more general dependence on the velocity of
the body relative to the fluid. The case of quadratic drag is substantially more
complex than the linear case, as it nonlinearly couples both components of the
momentum equation. Careful screening of the literature on this classical topic
showed that, unexpectedly, the solutions reported do not directly provide the
particle velocity as a function of time but use auxiliary quantities or apply to
special cases only. No explicit solution using
elementary operations on analytical expressions is known for a general
trajectory. After a detailed account of the literature, the paper provides such
a solution in form of a ratio of two series expansions. This result is
discussed in detail and related to previous approaches. In
particular, it is shown to yield, as limiting cases, certain approximate
solutions proposed in the literature. The solution technique employs a strategy to
reduce systems of ordinary differential equations with a triangular dependence
of the right-hand side on the vector of unknowns to a single equation in an
auxiliary variable. For the particular case of quadratic drag, the auxiliary
variable allows an interpretation in terms of canonical coordinates of motion
within the framework of Hamiltonian mechanics.
Another result of the paper is the extension of the solution technique to more
general drag laws, such as general power laws or power laws with an additional
linear contribution.
Furthermore, generalization to variable velocity of the surrounding fluid is
addressed by considering a linear velocity profile for which a solution is
provided as well. Throughout, the results obtained are illustrated by numerical
examples.

\end{abstract}

\section{Introduction}
\label{Sec:Intro}
Projectile motion constitutes a very elementary problem of classical mechanics;
as such, occupying a central place in the works of Niccol\`{o} Tartaglia,
Galileo Galilei and Sir Isaac Newton, to name but a few. Although Hardy
\cite{Hardy1940} scathingly remarked in his \emph{A Mathematician's Apology}
that the science of ballistics were ``repulsively ugly and intolerably dull'',
he still admitted that it does demand ``a quite elaborate technique''. The latter statement
perhaps explains why it was of interest to mathematicians such as Johann
Bernoulli, Leonhard Euler, Adrien-Marie Le Gendre and Johann Heinrich Lambert,
throughout the course of its long history.

In the present study, a projectile or any other object moving through a gas or
liquid is modelled as a point mass. Since the latter has no spatial extensions,
this amounts to neglecting issues of shape, orientation and rotation
altogether. On the other hand, the point mass used here is supposed to
experience a drag when moving through the surrounding medium, which obviously
results from the extension of the moving object. Furthermore, we suppose lift
forces to be negligible as compared to drag forces. The model considered here
is hence that of a point mass experiencing gravity and drag, and the terms
``projectile'', ``body'', ``object'', when used in the following, are meant in
this sense and used as synonyms. Furthermore, the term ``fluid'' is used to
encompass both liquids and gases. This model has been widely used in the
literature, such as the works cited subsequently. In line with those studies,
we assume the model to be a valid representation of physical reality to the
desired degree of accuracy and focus on providing mathematical solutions of the
model equations.

In Book II of his \emph{Philosophi{\ae} Naturalis Principia Mathematica}, Newton
\cite{Newton1687} gives the first rigorous treatment of this problem using
mathematical physics. A simple force balance can be used to account for the two
principal forces, gravity and drag, and using his laws of motion
(established in Book I of the same work), the equations of motion can be
formulated in vectorial notation to read:
\begin{equation}
m_{{\rm p}} \frac{{\rm d} {{\bf v}}}{{\rm d} t} = {{\bf F}}_{{\rm g}} +
{{\bf F}}_{{\rm d}}\;.
\label{eq:forcebalance}
\end{equation}

Here, \(m_{{\rm p}}\) denotes the mass of the moving object, \({\bf v}\) the
velocity vector of the object in a suitable inertial frame of reference \added{where} \({\bf
F}_{{\rm g}}~=~m_{{\rm p}}\,{\bf g}\) denotes the gravity force, with
\({\bf g}\) the constant gravity acceleration vector. A Cartesian coordinate
system \(\left(x,y\right)\) is usually chosen, with \(x\) the horizontal
coordinate and \(y\) the vertical coordinate, such that \({\bf g}\) is
oriented in the negative \(y\) direction, i.e. \({\bf g} =
\left(0,-g\right)\). The equations of motion then constitute a system of
ordinary differential equations in the components of \({\bf v} =
\left(v_x,v_y\right)\). \added{Furthermore} \({\bf F}_{\rm d}\) denotes the drag force acting
in the direction opposite to the velocity relative to the surrounding fluid.
The nature of the drag force then determines the nature of the equations of
motion and thereby the mathematical problem at hand. The statement of the
problem is completed by imposing the initial conditions \({\bf v}(t~=~0)~=~{\bf v}_0\)\added{, and the trajectory \({\bf x} = (x,y)\) may be found by integrating \({\bf v} = {\rm d}{\bf x}/{\rm d}t\) over time with initial conditions \({\bf x}(t=0) = {\bf x}_0\).}

Newton \cite{Newton1687} considered three forms of drag, namely linear,
quadratic and a superposition of the two. If the fluid resistance varies in
direct proportion with the relative velocity (i.e. linear or Stokes drag, valid
for low Reynolds numbers \({\rm Re} \ll 1\)), the differential equations for
\(v_x\) and \(v_y\) decouple due to the linearity of the drag force and
constitute a system of linear first-order ordinary differential equations. The
solution can be readily found using elementary methods (cf. textbooks on the
theory of ordinary differential equations, e.g. Walter \cite{Walter2000} \added{or Arnol'd \cite{Arnold1992}}). Should
the resistance, however, vary as the square of the velocity (quadratic drag,
valid for Reynolds numbers between \(10^3\) and \(10^5\), for example), one has
to deal with a system of nonlinearly coupled ordinary differential equations, and
the solution necessitates a more involved approach. Newton himself was unable
to solve the problem, but his contemporary, Johann Bernoulli, solved it after
being challenged by the British astronomer John Keill \cite{Bernoulli1719}.
Bernoulli's solution parameterizes the absolute value of the velocity over the
trajectory slope angle and is, hence, implicit. This solution is also known
as the \emph{hodograph solution}. The hodograph, being the locus of the tip of
the velocity vector with the other end held fixed, has precisely the two
afore-mentioned quantities as polar coordinates. Furthermore, the solution
contains quadratures which must be evaluated numerically. Despite these
drawbacks, it is the standard and by far most widely cited solution
\cite{Synge_Griffith1949,Hayen2003,Benacka2010,Benacka2011}.

Much of the literature on projectile motion with quadratic drag after Bernoulli
comprises efforts to calculate approximate solutions based on the hodograph
solution. Various exact and approximate implicit formul\ae{} using miscellaneous
series and approximation techniques can be found in the extensive literature
available on this subject, the most well-known of which are due to Euler
\cite{Euler1745}, Lambert \cite{Lambert1767}, Borda \cite{Borda1769} and Le
Gendre \cite{Legendre1782}.
A common feature of the cited works is the implicitness of the solutions derived
therein, in the following sense: The unknown quantities (the velocity
components \(v_x\; \textrm{and}\; v_y\) or alternatively the cooridinates of
the position of the particle \(x\;\textrm{and}\; y\)), instead of being
expressed as functions of time, are parameterized over some other auxillary
quantities, e.g. trajectory slope angle. Of particular interest is the
solution in \cite{Lambert1767}, where \(y\) and \(t\) are parameterized over
\(x\).
The exposition of their complete results is beyond the scope of the present paper \added{and} the
interested reader is referred to Isidore Didion's \emph{Trait\'{e} de
balistique}, sect.\ V, \textsection\, III--IV, where these results have been
painstakingly compiled \cite{Didion1860}. For a recent review on the historical
aspect of the problem, see also \cite{Hackborn2006}.

From the late 19th century onwards, numerical and empirical methods superseded
the more analytical approach of Bernoulli and Euler. Nevertheless, the problem
continues to afford scope for an instructive application of mathematical
analysis to physics. An approximate formula of the trajectory for low
quasi-horizontal paths is found in \cite{Lamb1923}. Parker \cite{Parker1977}
rederived the result and provided---to the best of our knowledge---for the
first time, approximate explicit solutions for flat trajectories and short
duration. He also arrived at the results of Bernoulli by following a
considerably different set of variable transformations. Another recent
approximation is due to Tsuboi \cite{Tsuboi1996}, who considered perturbative
solutions of the first order for small angles of release. Lastly, an explicit exact albeit
semi-analytical (in the sense that it requires recourse to numerical means)
solution was given by Yabushita {\itshape et al.} \cite{Yabushita2007} using the recently
developed homotopy analysis method developed by Liao \cite{Liao2004}.

In summary, it appears that in spite of the long history and the substantial
amount of material available on this subject, an exact explicit solution even
using analytic functions cannot, to the best of our knowledge, be found in the
existing literature. Providing such a solution would hence be of interest. In
the foregoing context, by ``exact" we mean a solution that satisfies the
complete equations of motion for the whole set of initial conditions of physical
interest, as opposed to requiring further approximations or simplifying
assumptions that are valid for a restricted set of initial conditions only. A
solution is said to be ``explicit" in the following if it depends explicitly on
the natural independent variable, here time, as opposed to being a function of
some other auxiliary variable (e.g. Bernoulli's hodograph solution \cite{Bernoulli1719},
where the solutions are implicitly expressed as functions of trajectory slope).
Finally, an ``analytic function" is defined in mathematical analysis as a
function which can locally be developed in terms of a convergent power series.
The latter is to be distinguished from a closed form expression which is given
by a finite set of elementary functions.

Drag laws more general than linear or quadratic have, for obvious reasons,
received significantly less attention. Bernoulli \cite{Bernoulli1719} provides an exact
implicit solution using the hodograph technique for cases where the drag force
varies as a general power of the velocity. Such laws are valid for high
velocity subsonic and some cases of supersonic projectile motion, with Reynolds
numbers substantially higher than \(10^5\) \cite{Weinacht+2005}.
For Newton's generalization involving a superposition of quadratic and linear
drag \cite{Newton1687}, no exact solutions can, to the best of our knowledge, be found
in the literature. According to Clift, Grace and Weber \cite{Clift1978}, it may be instructive
to consider even more general forms, especially for motion with Reynolds number
between \(1\;\textrm{and}\;10^3\). Although of no significance to exterior
ballistics, projectile motion in the transition regime between Newtonian and
Stokesian drag may be studied in order to gain general understanding of
particle motion in sediment transport phenomena---somewhat along the lines of
\cite{Nalpanis+1993}.

A final generalization considered in this paper deals with the effects of a
linear \added{fluid} velocity profile on projectile motion, as no studies investigating the
influence of a velocity profile of any kind on projectile motion with nonlinear
drag could be found in the literature.

The paper is structured as follows: We begin by establishing a useful principle
for the reduction of a certain class of nonlinear systems of ordinary
differential equations to a single equation. It is then exploited to solve the
basic problem of projectile motion with quadratic resistance. The various
inter-relationships between the solution thus obtained and the previous
attempts are then examined. Subsequently, the limiting behaviour of the
solution in cases of physical interest is investigated. An integral of motion
is derived, its relation to historical forms found in the literature is
elucidated and a physical interpretation is suggested based on the present
form.

In the next part, two possible generalizations of the basic problem are
proposed. First, the approach employed in solving the quadratic drag problem is
used to tackle more general drag laws. Thereafter, the scenario is generalized
by including the effect of a \added{fluid} velocity profile and the governing equations are
solved by virtue of the reduction principle established earlier. Finally, some
interesting properties of the solution are discussed.

\section{A reduction principle for certain coupled systems of ordinary differential equations}
\label{sec:RedPrinc}
Let \({\bf X}=\left(X_1,X_2,X_3,\ldots\right)\), be an \(n\)-dimensional
vector-valued function of \(t\in \mathbb{R}\) and let it be determined by the
initial value problem
\begin{equation}
\frac{{\rm d} {\bf X}}{{\rm d} t} + P{\bf X} = {\bf Q}\;, \qquad {\bf X}(t=0) = {\bf X}_0\;.
\label{eq:abstractodesimple}
\end{equation}

Here, \(P:\mathbb{R} \rightarrow \mathbb{R}\) is a scalar real-valued
function and \({\bf Q} =
\left(Q_1,Q_2,Q_3,\ldots\right)\colon~\mathbb{R}~\rightarrow~\mathbb{R}^n\) denotes a
vector-valued function. If \(P\) and \({\bf Q}\) were known functions depending only on \(t\), then the equations would constitute a linear system of ordinary differential equations and there would
be no difficulty in applying the method of variation of parameters \added{first proposed by Lagrange in \cite{Lagrange1809b} and discussed in more modern textbooks such as Arnol'd \cite{Arnold1992}} in order to write down the solution for
each and every component of \eqref{eq:abstractodesimple} in terms of \(P\), \({\bf Q}\) and appropriate quadratures involving the two, thus
\begin{equation}
{\bf X} = \frac{{{\bf X}}_{0} + \displaystyle\int_0^t{\!{\bf
Q}\exp\!\left(\displaystyle\int_0^\tau{\!P\,{\rm d} s}\right)}\,{\rm d} \tau}{\exp\!\left(\displaystyle\int_0^t{\!P\,{\rm d} \tau}\right)}\;.
\label{eq:abstractodesimplesol}
\end{equation}

For the purpose of this investigation, however, it is essential \added{to}
consider a more general class of equations. Assume now, that \(P\) is a scalar
function of \({\bf X}\) and (possibly) \(t\), i.e.\ \(P = P({\bf X};t)\).
Furthermore, assume that \({\bf Q} = {\bf Q}({\bf X};t)\), with the
constraint that \(\nabla_{\!{\bf X}}{\bf Q}\), which is a tensor of
rank two, be a strictly lower triangular matrix, i.e.
\begin{align}
Q_1 & = Q_1(t) \nonumber\\
Q_2 & = Q_2(X_{1};t) \nonumber\\
Q_3 & = Q_3(X_{1},X_{2};t) \nonumber\\
Q_4 & = Q_4(X_{1},X_{2},X_{3};t) \nonumber\\
& \;\;\vdots \nonumber
\end{align}
In that case, standard variation of parameters is inadequate and the expression
in \eqref{eq:abstractodesimplesol} can hardly be called a solution. Ignoring,
however, for a moment that the equations are no longer linear and naively using
variation of parameters on the first component of \eqref{eq:abstractodesimple},
the following formal expression is obtained:
\begin{equation}
X_1 = \frac{X_{1,0} + \displaystyle\int_0^t{\!Q_1(\tau)\exp\!\left(\displaystyle\int_0^\tau{\!P({\bf X};s)\,{\rm d} s}\right)}\,{\rm d} \tau}{\exp\!\left(\displaystyle\int_0^t{\!P({\bf X};\tau)\,{\rm d}\tau}\right)}\;.
\label{eq:abstractodetoughcomp1}
\end{equation}
The only unknown quantity is the exponential of \(\int P {\rm d} t\), and it will
later be clear that it is a key quantity in this method. Therefore, let
\begin{equation}
\phi \equiv \exp\!\left(\int_0^t P({\bf X};\tau)\,{\rm d}\tau\right)
\label{eq:abstractodedefphi}
\end{equation}
for the sake of brevity. Now, let us take the liberty of expressing
\eqref{eq:abstractodetoughcomp1} more succinctly by writing \(X_1 \equiv
f_1(\phi)\). This can be taken over to the next component of \({\bf X}\)
yielding
\begin{equation}
\eqalign{
X_{2} & = \frac{X_{2,0} + \displaystyle\int_0^t{\!Q_2(X_{1};\tau)\exp\!\left(\displaystyle\int_0^\tau{\!P({\bf X};s)\,{\rm d} s}\right)}\,{\rm d}\tau}{\exp\!\left(\displaystyle\int_0^t{P({\bf X};\tau)\,{\rm d}\tau}\right)} \cr
& = \frac{1}{\phi}\left(X_{2,0} + \displaystyle\int_0^t{\!Q_2\!\left(f_1(\phi);\tau\right)\phi\,{\rm d}\tau}\right)}
\end{equation}
This, again, is denoted as \(X_{2} \equiv f_2(\phi)\). Likewise, the \(i\)-th
component of \({\bf X}\) then may be expressed recursively in a similar
fashion. From a general point of view, this procedure defines an \(n\)-tuple of
mappings \({\bf f} = \left(f_1, f_2, f_3, \ldots\right): I \rightarrow D\)
from the space \(I\) of positive integrable functions to the space \(D\) of
differentiable functions, so that \(\eta \in I\) is mapped onto
\begin{equation}
\eqalign{f_1(\eta) & \equiv 
\frac{1}{\eta}\left(X_{1,0} + \displaystyle\int_0^t \!Q_1\!\left(\tau\right) \eta\,{\rm d}\tau\right) \label{eq:abstractodeconstruction} \cr
f_i(\eta) & \equiv \frac{1}{\eta}\left(X_{i,0} + \displaystyle\int_0^t \!Q_i\!\left(f_1(\eta),f_2(\eta),\ldots,f_{i-1}(\eta);\tau\right) \eta\,{\rm d} \tau\right) \qquad 1 < i \leqslant n}
\end{equation}
The solution of \eqref{eq:abstractodesimple} can now be formulated as
\begin{equation}
{\bf X} = {\bf f}(\phi) \label{eq:abstractodetoughsol}
\end{equation}
with \(\phi\) defined according to \eqref{eq:abstractodedefphi}. Inserting this
into the first component of \eqref{eq:abstractodesimple} gives
\begin{equation}
\frac{{\rm d} f_1(\phi)}{{\rm d} t} + P\!\left({\bf f}(\phi)\right)f_1(\phi) = Q_1(t)\;.
\label{eq:abstractodefinal}
\end{equation}
In this equation, the only unknown quantity involved is \(\phi\). Once \(\phi\)
is known, \({\bf X}\) is readily calculated using
\eqref{eq:abstractodeconstruction}--\eqref{eq:abstractodetoughsol}. Thus, the
original problem involving a non-linearly coupled system of ordinary
differential equations has been reduced to a single equation in an auxiliary
variable appropriately defined for that purpose. For ease of reference, we call
it the resolvent variable and the equation governing the auxiliary quantity the
resolvent equation.

The above formalism relieves one from the trouble of solving a whole set of coupled ordinary differential equations and rather allows one to concentrate the investigation on a single equation for a scalar quantity only. Since \({\bf f}\) involves quadratures
over (possibly) non-trivial kernels, the resolvent equation is, in general, an
integro-differential equation and, therefore, may not be trivial itself. It
clearly depends to a large extent on the nature of \(P\) and \({\bf Q}\) as
to how difficult the new equation is. It will, however, be seen in the
following that in the case of the particular class of problems considered here,
the formalism may be exploited to obtain elegant solutions of the respective
original systems of ordinary differential equations.

\added{While the basic idea of converting the problem of finding a solution to \eqref{eq:abstractodesimple} to solving \eqref{eq:abstractodefinal} is elementary and reminiscent of an elimination strategy for solving a linear algebraic system with a lower triagonal coefficient matrix, its use for the solution of a nonlinear system of ordinary differential equations appears to be new. Despite several efforts, the authors could not find this result in the available literature.}

\section{Projectile motion with quadratic resistance}
\label{sec:NewtonDrag}
\subsection{Mathematical formulation}
\noindent The general force balance \eqref{eq:forcebalance} discussed above is
now \added{completed} by specifying the drag force. The quadratic resistance law can be
written in the form
\begin{equation}
{\bf F}_{\rm d} = -\frac12 C_{\rm d}\varrho A\,\|{\bf v}-{\bf V}\|\left({\bf v}-{\bf V}\right)\;.
\end{equation}
Here, \(C_{\rm d}\) is the constant drag coefficient characteristic of the
projectile's shape, \(A\) is the cross-sectional area and \(\varrho\) is the
density of the fluid. The velocity of the fluid which constitutes the resistive
medium is denoted by \({\bf V}\), which, at present, \added{is} assume\added{d} to be constant
in space and time. Equation \eqref{eq:forcebalance}, therefore, reads
\begin{equation}
\frac{{\rm d} {\bf v}}{{\rm d} t} = -\alpha\,\|{\bf v}-{\bf V}\|\left({\bf v}-{\bf V}\right) + {\bf g}\;.
\label{eq:basicproblemodevec}
\end{equation}
where the constant \(\alpha\) is a shorthand for \(\frac12 C_{\rm d}\varrho
A/m_{\rm p}\). Finally, the quantity
\begin{equation}
{\bf u} = {\bf v} - {\bf V}
\label{eq:defRelVel}
\end{equation}
\added{is introduced}, which is the relative velocity between the particle and the fluid. Since \({\bf
V}\) is a constant vector, this yields
\begin{equation}
\frac{{\rm d} {\bf u}}{{\rm d} t} = -\alpha\,\|{\bf u}\| \, {\bf u} + {\bf g}
\label{eq:basicproblemodevecfinal}
\end{equation}
or, when using the components of \({\bf u} = \left(u_x, u_y\right)\)
\begin{align}
\frac{{\rm d} u_x}{{\rm d} t} + \alpha u_x\sqrt{u_x^2 + u_y^2} & = 0 \label{eq:basicproblemodexcomp}\\
\frac{{\rm d} u_y}{{\rm d} t} + \alpha u_y\sqrt{u_x^2 + u_y^2} & = -g\label{eq:basicproblemodeycomp}
\end{align}
Without loss of generality, the coordinate system may be placed at the starting
point of trajectory, \added{so that \(x_0 = y_0 = 0\)}. The initial conditions are given by \({\bf u}(0)={\bf
v}(0)-{\bf V}\), or, in components,
\begin{align}
u_x(0) & = v_{x,0}-V_{\!x} \equiv u_{x,0} \label{eq:basicproblemivxcomp} \\
u_y(0) & = v_{y,0}-V_{\!y} \equiv u_{y,0} \label{eq:basicproblemivycomp}
\end{align}
Observe that a purely vertical motion (in the reference frame moving at the
fluid velocity) is obtained if \(u_{x,0} = 0\). That case, again, is described by
one component only and the ordinary differential equation can be solved using
separation of variables. Here, however, we are interested in the more general
case and eliminate this situation by requiring that
\begin{equation}
u_{x,0} > 0\;.  \label{eq:initcond_ux}
\end{equation}
With this condition,
\eqref{eq:basicproblemodexcomp}--\eqref{eq:basicproblemivycomp} constitute the
initial value problem to be solved in the following.

\subsubsection{Reduction to a single scalar equation}
\paragraph{Review of Parker's approach} Since the present method of solution is
somewhat akin in spirit to Parker \cite{Parker1977}, it is
instructive to study the latter now. Division of \eqref{eq:basicproblemodeycomp} by
\eqref{eq:basicproblemodexcomp}, which is well-defined due to the condition
\eqref{eq:initcond_ux}, yields, after appropriate algebraic manipulation,
\begin{equation} 
\frac{{\rm d} u_y/{\rm d} t + g}{{\rm d} u_x/{\rm d} t} =
\frac{u_y}{u_x}\;.
\end{equation}
This can be simplified using the quotient rule of differential calculus and
subsequent separation of variables to give
\begin{equation}
u_y = u_x\left(\frac{u_{y,0}}{u_{x,0}}-g\int_0^t \frac{{\rm d}
\tau}{u_x}\right)\;.
\end{equation}
Inserting this into \eqref{eq:basicproblemodexcomp} and substituting 
\begin{equation}
w = g\int_0^t \frac{{\rm d} \tau}{u_x} - \frac{u_{y,0}}{u_{x,0}}
\end{equation}
yields the equation
\begin{equation}
\frac{{\rm d}^2w}{{\rm d} t^2}  = \alpha g \sqrt{1+w^2}\;.
\label{eq:basicproblemreducew}
\end{equation}
This is the resolvent equation \added{proposed in the cited reference}. The task of solving the
original system is now reduced to solving this scalar equation subject to the
initial conditions \(w(0) = -u_{y,0}/u_{x,0}\) and \({\rm d} w(0)/{\rm d} t = g/u_{x,0}\).
\paragraph{Modified approach} 
One may recast \eqref{eq:basicproblemodexcomp} and
\eqref{eq:basicproblemodeycomp} into the form considered in \secref{sec:RedPrinc}
by writing
\begin{equation}
P = \alpha\sqrt{u_x^2 + u_y^2} \qquad {\bf Q} = \left(0,-g\right)\;.
\label{eq:PQ_quadDrag}
\end{equation}
Since \(\nabla_{\!{\bf u}}{\bf Q}\) is a null matrix, the
condition of applicability is trivially fulfilled. It is, therefore, natural to
chose
\begin{equation}
\phi \equiv \exp\!\left(\alpha\!\displaystyle\int_0^t{\!\!\sqrt{u_x^2 + u_y^2}\,{\rm d} \tau}\right)
\end{equation}
in order to express \({\bf u}\) as
\begin{align}
u_x & = \frac{u_{x,0}}{\phi} \label{eq:basicproblemxcompform}\\
u_y & = \frac{1}{\phi}\left(u_{y,0}-g\displaystyle\int_0^t\!\phi\,{\rm d}
\tau\right) \label{eq:basicproblemycompform}
\end{align}
Inserting \added{this} into \eqref{eq:basicproblemodexcomp} yields
\begin{equation}
\frac{{\rm d} \phi}{{\rm d} t} = \alpha \sqrt{u_{x,0}^2 + \left(u_{y,0}-g\displaystyle\int_0^t\!\phi\,{\rm d} \tau\right)^{\!2}}\;,
\end{equation}
which is an integro-differential equation. However, since the
only kernel involved is unity itself, the equation can be reduced to an
ordinary differential equation by rewriting it in terms of the quantity
\begin{equation}
\Phi \equiv \int_0^t \phi \,{\rm d} \tau
\label{eq:order_reductio}
\end{equation}
to yield
\begin{equation}
\frac{{\rm d}^2\Phi}{{\rm d} t^2} = \alpha \sqrt{u_{x,0}^2 + \left(u_{y,0}-g\,\Phi\right)^{2}}\;.
\label{eq:basicproblemreducephi}
\end{equation}
The initial conditions follow directly from the definition of the quantities
\(\Phi\) and \(\phi\) as \(\Phi(0) = 0\) and \({\rm d} \Phi(0)/{\rm d} t = \phi(0) = 1\).
This is the new initial value problem equivalent to the original system which
will be investigated in the subsequent part of the paper.

\paragraph{Remark} The result of Parker's approach and the present one are
essentially equivalent, which is readily shown by observing that \(\,w =
\left(g\,\Phi - u_{y,0}\right)/u_{x,0}\). The difference between Parker's
approach and the present one resides rather in the method than in the result.
The present method can be readily generalized to tackle more involved problems,
such as the one \added{accounting for fluid} velocity profiles, as shown later, while
Parker's method is not so flexible. To be precise, the approach from
\cite{Parker1977} is only applicable as long as \(\nabla_{{\bf u}}{\bf Q}\) is a null
matrix, whereas for the present method, it only needs to be a strictly lower
triangular matrix. Therefore, this formalism may also be regarded as a means of
embedding the solution techniques employed for the present problem into a more
general framework.

\subsubsection{Solution of the reduced problem}
\noindent Since the right hand side of \eqref{eq:basicproblemreducephi} can be
developed in a power series around the origin (\(t = 0\)), the classical theory of ordinary
differential equations \cite{Walter2000} then guarantees the existence of a power
series expansion for \(\Phi\) around \(t = 0\), i.e. \(\Phi = a_0 + a_1 t + a_2
t^2 + a_3 t^3 + \cdots\) with
\begin{equation}
a_j = \frac{1}{j!}\frac{{\rm d} ^j \Phi(0)}{{\rm d} t^j}\;.
\end{equation} 
according to Taylor's theorem. The first two coefficients are determined using
the initial conditions yielding \(a_0 = 0\,\textrm{and}\,a_1 = 1\). Evaluating
\eqref{eq:basicproblemreducephi} at \(t = 0\) provides a direct relation for the
third coefficient:
\begin{equation}
a_2 = \frac{\alpha \sqrt{u_{x,0}^2+\left(u_{y,0} - ga_0\right)^2}}{2} = \frac{\alpha \sqrt{u_{x,0}^2 + u_{y,0}^2}}{2}\;.
\end{equation}
Differentiating \eqref{eq:basicproblemreducephi} once at \(t = 0\) and algebraic
rearrangement gives
\begin{equation}
a_3 = -\,\frac{\alpha g u_{y,0}}{6 \sqrt{u_{x,0}^2 + u_{y,0}^2}}\;.
\end{equation}
This process, continued \emph{ad infinitum}, yields all the coeffients of the
series. In general, differentiating both sides of Eq.
\eqref{eq:basicproblemreducephi} \(n\) times with respect to \(t\) gives
\begin{equation}
a_{j+2} = \frac{\alpha}{\left(j+2\right)!}\left.\frac{{\rm d}^{j}}{{\rm d} t^{j}}\sqrt{u_{x,0}^2 + \left(u_{y,0}-g\,\Phi\right)^2}\right|_{t=0}.
\label{eq:basicproblempowerseriesphi}
\end{equation}
This, in conjunction with the initial conditions, constitutes a recursive
formula for the coefficients of the power series expansion of \(\Phi\), which
solves the problem. For the sake of convenience, let \(\,u_0 \equiv \|{\bf u}_0
\| = \sqrt{u_{x,0}^2 + u_{y,0}^2}\) be the Euclidean norm of the initial
velocity vector and let \(\theta_0 \equiv \arctan
\left(u_{y,0}\middle/u_{x,0}\right)\) be the initial angle of the trajectory.
Using \(u_{x} = \left.u_{x,0}\middle/({\rm d} \Phi/{\rm d} t)\right.\) and \(u_{y} =
\left(u_{y,0} - g\,\Phi\right)/({\rm d} \Phi/{\rm d} t)\), the final solution is
\begin{equation}
\eqalign{
\Phi & = t + \frac{\alpha u_0}{1\cdot 2} \, t^2 - \frac{\alpha g \sin \theta_0}{1\cdot 2\cdot 3} \, t^3\cr
& \quad \,\,\,\, + \frac{\alpha g^2 u_{0}^{-1} \cos^2 \!\theta_0 - \alpha^2 g u_0 \sin \theta_0}{1\cdot 2\cdot 3\cdot 4} \, t^4 \cr
& \quad \,\,\,\, + \,\cdots \label{eq:Phiterms}
}
\end{equation}
While the existence of a solution of this type for
\eqref{eq:basicproblemreducephi}, i.e. in the form of such a series expansion,
results from classical theory, this series so far has never been provided and
investigated per se. According to \eqref{eq:basicproblemxcompform} and
\eqref{eq:basicproblemycompform}, one subsequently obtains the solution for
\(u_x,u_y\) as the ratio of two power series, whereby the numerator of \(u_x\)
degenerates to a constant. Again, we have reason to believe (cf.
\secref{Sec:Intro}) that a solution of this form cannot be found in the
literature, thus constituting one of the contributions of the present work.
\begin{align}
u_x & = \frac{u_{x,0}}{1 + \alpha u_0 \, t - \displaystyle\frac{\alpha g \sin \theta_0}{1\cdot 2} \, t^2 + \displaystyle\frac{\alpha g^2 u_0^{-1} \cos^2 \!\theta_0 - \alpha^2 g u_0 \sin \theta_0}{1\cdot 2\cdot 3} \, t^3 + \,\cdots} \label{eq:u_x-terms} \\
u_y & = \displaystyle\frac{u_{y,0}}{1 + \alpha u_0 \, t - \displaystyle\frac{\alpha g \sin \theta_0}{1\cdot 2} \, t^2 + \displaystyle\frac{\alpha g^2 u_0^{-1} \cos^2 \!\theta_0 - \alpha^2 g u_0 \sin \theta_0}{1\cdot 2\cdot 3} \, t^3 + \,\cdots} \label{eq:u_y-terms}\\
\quad & \quad - g\,t \, \frac{1 + \displaystyle\frac{\alpha u_0}{1\cdot
2} \, t - \displaystyle\frac{\alpha g \sin \theta_0}{1\cdot 2\cdot 3} \, t^2 +
\displaystyle\frac{\alpha g^2 u_0^{-1} \cos^2 \!\theta_0 - \alpha^2 g u_0 \sin
\theta_0}{1\cdot 2\cdot 3\cdot 4} \, t^3 + \,\cdots}{1 + \alpha u_0 \, t -
\displaystyle\frac{\alpha g \sin \theta_0}{1\cdot 2} \, t^2 +
\displaystyle\frac{\alpha g^2 u_0^{-1} \cos^2 \!\theta_0 - \alpha^2 g u_0 \sin
\theta_0}{1\cdot 2\cdot 3} \, t^3 + \,\cdots} \nonumber
\end{align}
Finally, one can return to the corresponding absolute quantities with \(\,v_x =
u_x + V_x\,\,{\rm and}\,\,v_y = u_y + V_y\).

\added{Providing a function in terms of a power series, or a ratio of power series, raises the question of radius of convergence. We have made several attempts to solve this issue by applying various techniques. Unfortunately, it was so far not possible to analytically determine such an expression for the convergence radius from the recursion formula \eqref{eq:basicproblempowerseriesphi}, so that this is still an open problem. In the numerical test cases, a selection of which is provided below, no problems of convergence were encountered and the evaluation of partial sums was well-behaved. Details are reported in \secref{sec:num_illustr} below.}

\subsubsection{Relation of the present solution to previous results}
\paragraph{Comment on Parker's implicit solution} Parker \cite{Parker1977} integrated
\eqref{eq:basicproblemreducew} once to obtain
\begin{equation}
\frac{{\rm d} w}{{\rm d} t} = \sqrt{{\rm C}-\alpha g\left(w\sqrt{1+w^2} + \arsinh
w\right)}\;,
\end{equation}
where \({\rm C}\) is a constant completely determined by the initial conditions.
Separation of variables yields
\begin{equation}
t = \int_{-u_{y,0}/u_{x,0}}^w\frac{{\rm d} \omega}{\sqrt{{\rm C}-\alpha g\left(\omega\sqrt{1+\omega^2} + \arsinh \omega\right)}}\;.
\label{eq:parkermess}
\end{equation}
Perhaps in part due to the daunting nature of the integrand, he remarked that
``even if the indefinite integral could be evaluated, we would not be able to
invert the resulting expression to obtain an explicit solution''\added{, and stopped at this stage.} While it is
probably \added{true} that the integral cannot be evaluated without resorting to
numerical quadrature, \added{we found} that due to a recent result
from \cite{Dominici2003}, it is now possible to obtain an explicit solution in the
form of a power series from \eqref{eq:parkermess} without having to evaluate the
integral in the first place. The main result of the cited reference states that
if a function \(Y\) is implicitly given as
\begin{equation}
X = \int_0^Y \frac{{\rm d} Y^\prime}{f}\;,
\end{equation}
the relation can be solved for \(Y\) and has the form
\begin{equation}
Y = Y(X=0) + f(Y\!=0)\sum_{n=0}^\infty \left\{ \mathfrak{D}^n[f](Y\!=0)\frac{X^{n+1}}{(n+1)!}\right\}\;,
\end{equation}
where \(\mathfrak{D}^n[f](Y)\) denotes the \(n\)-th nested derivative of \(f\)
with respect to \(Y\), defined recursively as
\begin{align}
\mathfrak{D}^0[f](Y) & = 1 \\
\mathfrak{D}^{n+1}[f](Y) & = \frac{{\rm d}}{{\rm d}
Y}\left\{f\cdot\mathfrak{D}^n[f](Y)\right\}
\end{align}
Writing, for the sake of convenience, \(\dot{w} = {\rm d} w/{\rm d} t\) and
applying Dominici's theorem yields
\begin{equation}
w = -\,\frac{u_{y,0}}{u_{x,0}} + \frac{g}{u_{x,0}}\sum_{n=0}^\infty
\mathfrak{D}^n[\dot{w}](w=-\!\left.u_{y,0}\middle/u_{x,0}\right.)\frac{t^{n+1}}{(n+1)!}\;.
\end{equation}
Due to the fact that \(w = \left(g\,\Phi - u_{y,0}\right)/u_{x,0}\), the
following alternative expression for \(\Phi\) is derived:
\begin{equation}
\Phi =
\sum_{n=0}^\infty\left\{\frac{\mathfrak{D}^n[\dot{w}](w=-\tan
\theta_0)}{n!}\,\frac{t^{n+1}}{n+1}\right\}\;.
\end{equation}
Calculating the individual terms of this series and comparing them with the
ones predicted by \eqref{eq:Phiterms} shows that the two expressions are identical.

\paragraph{Relation to the hodograph solution}
The hodograph solution due to Johann Bernoulli \cite{Bernoulli1719} reduces the original
system to a single ordinary differential equation with \(u \equiv \| {\bf u} \|
= \sqrt{u_x^2 + u_y^2}\), the Euclidean norm of the relative velocity vector,
as dependent variable and \(\theta\) defined by \(\tan \theta \equiv {\rm d}
y^\prime/{\rm d} x^\prime\) as the independent variable. Bernoulli's approach can
be recovered from the present one as follows. Dividing
\eqref{eq:basicproblemycompform} by \eqref{eq:basicproblemxcompform}, rearranging
and using parametric differentiation yields the identity
\begin{equation}
\frac{u_y}{u_x} = \frac{u_{y,0}-g\,\Phi}{u_{x,0}} = \tan \theta\;.
\label{eq:bernoullimesspre1}
\end{equation}
Together with \eqref{eq:basicproblemxcompform}, this yields
\begin{equation}
u = \frac{\sqrt{u_{x,0}^2 + \left(u_{y,0}-g\,\Phi\right)^2}}{\phi}\;.
\label{eq:bernoullimesspre2}
\end{equation}
This relation, along with the identity
\begin{equation}
\frac{{\rm d}^2\Phi}{{\rm d} t^2} = \frac{{\rm d}}{{\rm d} t}\!\left(\frac{{\rm
d} \Phi}{{\rm d} t}\right) = \frac{{\rm d} \Phi}{{\rm d} t}\,\frac{{\rm d}
}{{\rm d} \Phi}\!\left(\frac{{\rm d} \Phi}{{\rm d} t}\right) = \phi\,\frac{{\rm
d} \phi}{{\rm d} \Phi}
\label{eq:phasetransform}
\end{equation}
and an elementary (but cumbersome) application of the chain rule of
differential calculus yields
\begin{equation}
\frac{{\rm d} u}{{\rm d} \theta} = u\tan \theta + \frac{\alpha u^3}{g \cos \theta}\;,
\end{equation}
a so-called Bernoulli differential equation (named after Jakob Bernoulli) for
\(u\) in terms of \(\theta\). Although the auxiliary equation here has an
analytic solution which can, in fact, be expressed in terms of elementary
functions, in order to change from the parameterization by \(\theta\) to the
one by \(t\), quadratures are necessary which cannot be evaluated without
recourse to numerics. This limits the usefulness of the approach.
Nevertheless, the fact that one approach can be recovered from the other may be
regarded as a consistency check.

\subsection{Properties}
\noindent \added{This section} illustrate\added{s} \added{the kind of} information \added{that} may be extracted from the
present solution \eqref{eq:Phiterms}--\eqref{eq:u_y-terms}, with the ultimate goal of
obtaining further insight into the properties of the motion.

\subsubsection{Integral of motion}
\noindent Contrary to physical intuition, projectile motion under quadratic drag
obeys a conservation law. This may be seen when using \eqref{eq:phasetransform} to
rewrite \eqref{eq:basicproblemreducephi} as
\begin{equation}
\phi \, {\rm d} \phi - \alpha \,  \sqrt{u_{x,0}^2 + \left(u_{y,0} - g \Phi\right)^2}\,{\rm d} \Phi = 0\;,
\end{equation}
which is readily integrated to yield
\begin{equation}
\textstyle\frac{1}{2}\,\phi^2 + U_*(\Phi) = {\rm const.}
\label{eq:Energy_mass}
\end{equation}
Here, $U_*(\Phi) \equiv - \alpha\int\!\!\sqrt{u_{x,0}^2 +
(u_{y,0} - g\,\Phi)^2}\,{\rm d}\Phi$. The quadrature
involved is elementary. An earlier and more common form for such an integral of motion is \cite{Didion1860}
\begin{equation}
-\,\frac{1}{2}\,\frac{1}{v^2\cos^2 \theta} + \xi(\theta) = {\rm
const.} \label{eq:didion1860}
\end{equation}
where $\xi(\theta) \equiv \int\!\left|\sec^3 \theta\right|\,{\rm d}
\theta$ (up to constant factors of historical nature). Although the two forms are equivalent,
\eqref{eq:Energy_mass} may have certain advantages insofar as physical
interpretation is concerned. Indeed, multiplying \eqref{eq:Energy_mass} with
$\mu_{\rm p} = \frac{1}{2}m_{\rm p}u_0^2$ and defining $\mu_{\rm p}U_* \equiv U$, one obtains
\begin{equation}
\textstyle\frac{1}{2}\mu_{\rm p}\phi^2 + U(\Phi) = {\rm const.}
\end{equation}
In terms of analytical mechanics, one may declare \(q
\equiv \Phi\) the canonical or Darboux coordinate of the motion and $p \equiv
\mu_{\rm p} \phi$ the conjugate momentum with the standard Poisson bracket
operator denoted $\{\cdot,\cdot\}$. The Hamiltonian of the system may then be
formulated as
\begin{equation}
\mathcal {H}(q,p) = \frac{p^2}{2\mu_{\rm p}} + U(q)\;.
\end{equation}
Further application of the chain rule of differential calculus shows that the
system of equations obtained upon applying the time evolution operator of
Hamiltonian mechanics
\begin{equation}
\frac{{\rm d} q}{{\rm d} t} = - \{\mathcal {H},q\} \qquad \frac{{\rm d} p}{{\rm
d} t} = - \{\mathcal {H},p\}
\end{equation}
is indeed equivalent to the original equations of motion derived from the
force balance using Newtonian mechanics. This nice relation to the Hamiltonian
formalism is not apparent when working with the hodographic quantities $u$ and
$\theta$.

\subsubsection{Limiting behaviour}
\label{sec:NewtonDrag:subsec:Prop:subsubsec:LimBeh}

\paragraph{Flat trajectories} A well-known approximate closed-form solution
valid for quasi-horizontal trajectories with \(\tan \theta \ll 1\) is due to
Lamb \cite{Lamb1923}. The condition is equivalent to \(u_y \ll u_x\) and necessarily
requires \(u_{y,0} \ll u_{x,0}\). Furthermore, it is required that the observed
time interval be sufficiently short. This is clear\added{ly} physical, since some time
or the other, the projectiles trajectory will turn downwards and become
steeper.\footnote{Parker \cite{Parker1977} derived this rigorously by requiring that
\(u_x^2 \gg u_y^2\), so that the approximation \(u \approx u_x\) may be made.
Then the equations of motion decouple and can, in fact, be solved in closed
form. Imposing the same condition on the solutions yields the requirements
\(u_{y,0}^2/u_{x,0}^2 \ll 1\) and \(t^2 \ll \alpha g\).} \added{It can be shown that under these prerequisites, the solution for the velocity can be found by elementary means in closed form and is given by} 
\begin{equation}
u_x = \frac{u_{x,0}}{1 + \alpha u_{x,0} t} \qquad \qquad u_y = \frac{u_{y,0} -
g\left(t + \frac12 \alpha u_{x,0} t^2\right)}{1 + \alpha u_{x,0} t}\;.
\label{eq:Parker}
\end{equation}

\added{On the other hand, truncating the series for \(\Phi\) in \eqref{eq:Phiterms} after the second-order term (so that the lowest neglected order in \(\phi = {\rm d}\Phi/{\rm d}t\) is second-order in \(t\)) yields}
\begin{equation}
\Phi \approx t + \textstyle\frac12 \alpha u_{x,0} t^2\;,
\end{equation}
since \(u_0 \approx u_{x,0}\) for \added{small} angles of release. This \added{turns equations \eqref{eq:u_x-terms},\eqref{eq:u_y-terms} into the expressions in \eqref{eq:Parker}.}
Setting \({\bf V} = 0\) in \eqref{eq:defRelVel}, which is tacitly assumed in
both \cite{Lamb1923} and \cite{Parker1977}, one obtains upon integration \added{over time} and
elimination of \(t\) \added{the trajectory}
\begin{equation}
y = \left(\frac{u_{y,0}}{u_{x,0}} + \frac{g}{2\alpha u_{x,0}^2}\right)x - \frac{g}{4\alpha^2 u_{x,0}^2}\left({\rm e}^{2\alpha x} - 1\right)\;,
\label{eq:quasi-horizontal}
\end{equation}
which is the well-known solution for flat trajectories.

\paragraph{Weakly resistive media} Often, one is interested in media where the
effect of resistance is still non-negligible, but not \added{too}
pronounced, so that \(\alpha\) may be taken to be small. In such cases,
perturbative techniques may be used to obtain approximate solutions. One such
result \added{obtained} by Tsuboi \cite{Tsuboi1996} can be rederived from the present
solution. The \added{approach} is based on a first-order perturbation in the small
parameter \(\alpha\) and a polynomial expansion in \(t\); third-order for
\(u_x\) and fourth-order for \(u_y\) (the degree of the \(u_y\) approximate can
be shown to be a necessary consequence of terminating the expression for
\(u_x\) at the third-order term in \(t\)). Therefore, the power series in the
denominator of the rational expression of \(u_x\) in \eqref{eq:u_x-terms} is
terminated at the third-order term, neglecting quadratic and higher-order
contributions in \(\alpha\), to obtain
\begin{equation}
u_x \approx \frac{u_{x,0}}{1 + \alpha u_0 t - \frac{1}{2}\alpha g \sin (\theta_0) t^2 + \frac{1}{6}\alpha g^2 u_0^{-1}\cos^2 (\theta_0) t^3}\,.
\end{equation}
This expression can now be formally manipulated using the Cauchy product to
yield
\begin{equation}
u_x \approx u_{x,0} - \alpha t\left(\frac{g^2 u_{x,0}^3}{6u_0^3} t^2 - \frac{g u_{x,0} u_{y,0}}{2u_0} t + u_{x,0}u_{0}\right),
\end{equation}
where again all contributions involving higher powers of \(\alpha\) have been
neglected. Finally, Tsuboi \cite{Tsuboi1996} made the assumption that \(u_{x,0} \gg
u_{y,0}\) or \(u_{0} \approx u_{x,0}\) (i.e. low take-off angle), so that
\begin{equation}
u_x \approx u_{x,0} - \alpha t\left[\textstyle\frac16 g^2 t^2 - \textstyle\frac12 u_{y,0} g t + \left(2u_{x,0}^2 + u_{y,0}^2\right)\right] \label{eq:tsuboiux}.
\end{equation}
Since the power series for \(\phi\) was terminated at the third-order term in
the expression of \(u_x\), for the sake of consistency, the same should be done
for \(u_y\), so that \added{the series for} \(\Phi = \int\!\phi\,{\rm d}t\) \added{has} to be terminated at the fourth-order
term. Using Cauchy products and
neglecting contributions from terms \added{of} higher order in \(\alpha\) yields
\begin{equation}
\eqalign{
u_y & \approx u_{y,0} - gt + \frac{\alpha t}{u_{x,0}}\Bigl[\textstyle\frac18 g^3 t^3 - \textstyle\frac12 u_{y,0}g^2 t^2 \cr & \qquad \qquad \qquad \qquad \left. + \textstyle\frac14 \left(2u_{x,0}^2 + 3u_{y,0}^2\right)g t - \textstyle\frac12 u_{y,0}\left(2u_{x,0}^2 + u_{y,0}^2\right)\right]. \label{eq:tsuboiuy}}
\end{equation}
Expressions \eqref{eq:tsuboiux} and \eqref{eq:tsuboiuy} are in agreement with the
result of \cite{Tsuboi1996} and illustrate that this can be seen as a special case
of the general solution presented in this paper.

\subsection{Numerical illustration}
\label{sec:num_illustr}
\begin{figure}
\centering
\includegraphics[width = 0.65\textwidth]{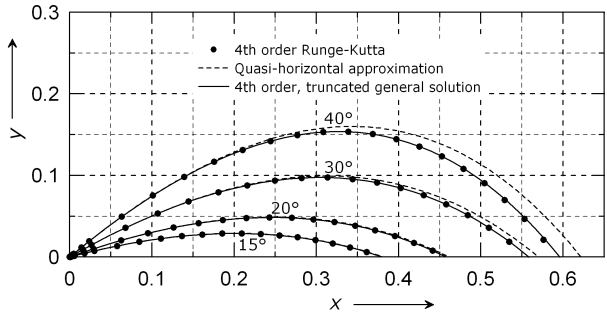}
\caption{Trajectories for the motion of a particle under quadratic drag in a stationary medium. The units are chosen such that $\alpha = g = 1$. The initial conditions are $v_0 = 1$ and (a) $\theta_0 = 15^\circ$, (b) $\theta_0 = 20^\circ$, (c) $\theta_0 = 30^\circ$ and (d) $\theta_0 = 40^\circ$. The 4th order Runge-Kutta solution of the initial value problem \eqref{eq:basicproblemodexcomp}--\eqref{eq:basicproblemivycomp} is compared with the quasi-horizontal approximation \eqref{eq:quasi-horizontal} and the fourth-order solution obtained from the general solution \eqref{eq:Phiterms}--\eqref{eq:u_y-terms}.}
\label{fig:ProjMQuadDrag2}
\end{figure}

\noindent For the following numerical examples, physical units are chosen such
that \(\alpha = g = 1\), which may be achieved by chosing the terminal velocity
\(v_\infty = \sqrt{g/\alpha}\) as the characteristic velocity and
\(v_\infty/g\) as the characteristic time scale. Also, we restrict ourselves to
stationary media, i.e. \({\bf V} = 0\) or \({\bf v} = {\bf u}\). In
\figref{fig:ProjMQuadDrag2}, a projectile is launched at various angles and with
initial velocity \(v_0 = 1\), mirroring approximately the choice of
\cite{Parker1977}.
For low initial angles (\(\theta_0 \lessapprox 20^\circ\)), the quasi-horizontal
solution \eqref{eq:quasi-horizontal} agrees well with the numerical solution of
the full problem obtained from a fourth-order Runge-Kutta method. For the sake
of consistency, it may be noted that using higher-order terms in the
approximation does not result in a noticeable difference. For a higher angle of
release, such as \(30^\circ\), the shortcomings of \eqref{eq:quasi-horizontal},
which is accurate only up to the first order in \(t\), as discussed in
\secref{sec:NewtonDrag:subsec:Prop:subsubsec:LimBeh}, become apparent. It
begins to deviate from the actual trajectory by considerable margins. On the
other hand, using higher-order terms, i.e.
\eqref{eq:Phiterms}--\eqref{eq:u_y-terms}, leads to a better approximation,
distinctly so for \(\theta_0 = 40^\circ\) \added{e.g.}, where the assumption of a fairly
flat trajectory is clearly violated.

\begin{figure}
\centering
\includegraphics[width = 0.65\textwidth]{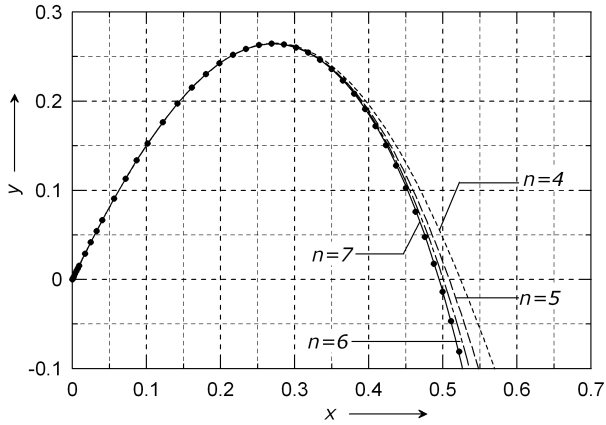}
\caption{Trajectory of an object released at a velocity $v_0 = 1$ and an initial angle of $\theta_0 = 60^\circ$ in units such that $\alpha = g = 1$. The numerical solution of Eqs. \eqref{eq:basicproblemodexcomp}--\eqref{eq:basicproblemivycomp} by a 4th order Runge-Kutta method (\textbullet) is compared with the solution obtained from \eqref{eq:u_x-terms}--\eqref{eq:u_y-terms} upon truncation of $\Phi$ after the term of order $n = 4,5,6,7$.}
\label{fig:ProjMQuadDrag60Grad}
\end{figure}

\begin{figure}
\centering
\includegraphics[width = 0.65\textwidth]{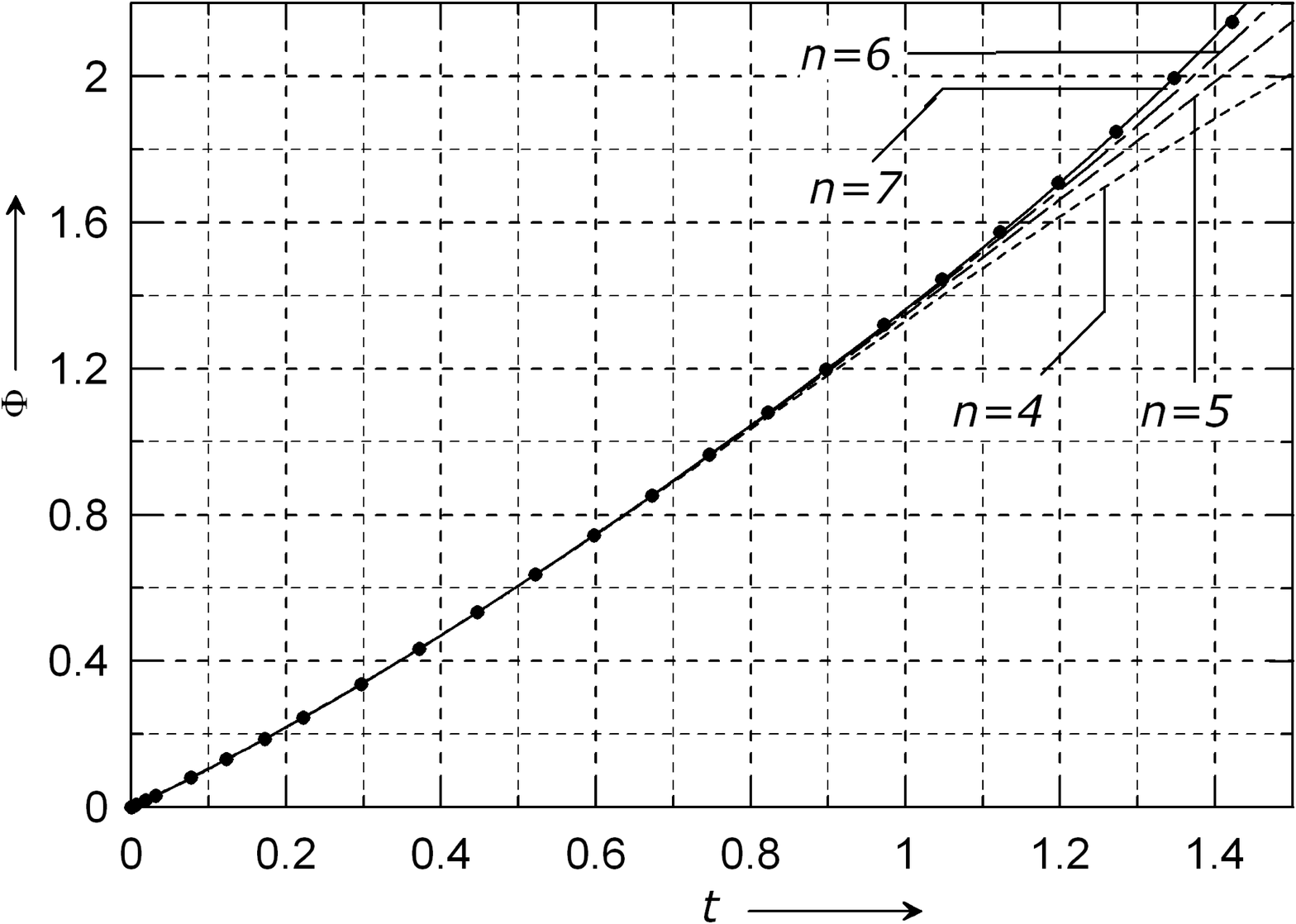}
\caption{The resolvent variable $\Phi$ of an object released at $v_0 = 1$ and initial angle $\theta_0 = 60^\circ$ as a function of time. The numerical solution of \eqref{eq:basicproblemodexcomp}--\eqref{eq:basicproblemivycomp} by a 4th order Runge-Kutta method (\textbullet) is compared with the solution obtained from \eqref{eq:u_x-terms}--\eqref{eq:u_y-terms} upon truncation of $\Phi$ after the term of order $n = 4,5,6,7$.}
\label{fig:ProjMQuadDrag60GradPhi}
\end{figure}

In \figref{fig:ProjMQuadDrag60Grad}, the trajectory of an object released at
\(v_0 = 1, \theta_0 = 60^\circ\) is depicted. Comparison is made between the
numerical solution and the solution obtained from
\eqref{eq:u_x-terms}--\eqref{eq:u_y-terms} upon truncation of the power series
for \(\Phi\) at various orders \(n\). For the sake of clarity, lower-order
approximations are not shown in the figure. In
\figref{fig:ProjMQuadDrag60GradPhi}, the result obtained numerically from
solving \eqref{eq:basicproblemreducephi} with a Runge-Kutta method is compared
with the partial sums of the series for $\Phi$, i.e. \eqref{eq:Phiterms}. The
lower order approximations are once again omitted in order to avoid unnecessary
cluttering.
From both figures, it is apparent that including higher order terms leads to an
increase in accuracy. For a solution based on power series, this is generally a
suggestive indication of convergence.


\section{Generalizations}
\label{sec:Gen}
In this section, two possible generalizations of the basic problem are
discussed. First, more general drag laws are considered. \added{Second},
the requirement of constant fluid velocity is relaxed in order to \added{include fluid velocity profiles which} depend on the vertical coordinate.
\subsection{Generalized Drag Laws}
\label{sec:Gen:subsec:Drag}
\noindent Here, the drag force is assumed to be of the form
\begin{equation}
{\bf F}_{\rm d} = -m_{\rm p}\left(\alpha\| {\bf v}-{\bf V}\|^{2k} +
\beta\right)\left({\bf v}-{\bf V}\right)\,,
\label{eq:gendrag}
\end{equation}
where $\alpha, \beta$ and $k$ denote real non-negative constants. The reason
this form of drag presents such an interesting object of study is two-fold.
First, from a mathematical point of view, this law is a formal generalization
of several simpler laws. Observe that setting $\alpha = 0$ reduces
\eqref{eq:gendrag} to the well-known linear Stokesian drag. For $\beta=0$, one
recovers the general power law, and further setting $k=1/2$ yields Newtonian
drag as used in \eqref{eq:basicproblemodevec}. Finally, arbitrary $\alpha,\beta$
and $k = 1/2$ corresponds to the generalization of the quadratic drag law
proposed by Newton \cite{Newton1687}. From a physical perspective, the form of
\eqref{eq:gendrag} comprises as a special case an often used correlation for the
drag force of a sphere proposed by Schiller and Naumann
\cite{SchillerNaumann1933}.\footnote{More precisely, Schiller-Naumann drag is
usually stated in the form \(\| {\bf F}_{\rm d} \| = \frac{1}{2}C_{\rm d} A
\| {\bf v} - {\bf V} \|^2\) with \(C_{\rm d} = C_{\rm d}({\rm Re}) =
24/{\rm Re}\,\left(1 + 0.15\,{\rm Re}^{0.687}\right)\) where ${\rm Re} =
\|{\bf v} - {\bf V}\| d/\nu$ denotes the particle Reynolds number with
\(d\) the particle diameter and \(\nu\) the kinematic viscosity of the fluid.} \added{In} \cite{Clift1978} \added{it is} reported that the Schiller-Naumann correlation is
accurate to within $5\%$ for Reynolds numbers between $0.2$ to $10^3$. This
drag law hence bridges the grey area between the Stokesian and the Newtonian
regime.

\subsubsection{Mathematical formulation}
\noindent Inserting \eqref{eq:gendrag} into the general force balance
\eqref{eq:forcebalance} yields
\begin{equation}
\frac{{\rm d} {\bf v}}{{\rm d} t} = -\left(\alpha\| {\bf v}-{\bf V}\|^{2k} + \beta\right)\left({\bf v}-{\bf V}\right) + {\bf g}\;.
\end{equation}
Rewriting the above in terms of the relative velocity \({\bf u} = {\bf v}-{\bf
V}\) and noting that \({\bf V} = {\rm const.}\) yields
\begin{equation}
\frac{{\rm d} {\bf u}}{{\rm d} t} = -\left(\alpha\| {\bf u}\|^{2k} + \beta\right){\bf u} + {\bf g}\;.
\end{equation}
In a Cartesian coordinate system identical to the one used in the previous
sections, this equation may be written in components as
\begin{eqnarray}
\frac{{\rm d} u_x}{{\rm d} t} + \left[\alpha\left(u_x^2+u_y^2\right)^k + \beta\right]u_x & = 0 \\
\frac{{\rm d} u_y}{{\rm d} t} + \left[\alpha\left(u_x^2+u_y^2\right)^k + \beta\right]u_y & = -g
\end{eqnarray}
The initial conditions are \({\bf u}(0) = \left(u_{x,0},u_{y,0}\right)\).

\subsubsection{Solution}
\label{subsubsec:solutiongendrag}
\noindent Let us begin by recasting the equations of motion into the form
considered in \secref{sec:RedPrinc}, setting \(P =
\alpha(u_x^2+u_y^2)^k + \beta\) and \({\bf Q} =
\left(0,-g\right)\).
Observe that, as in the previous problem, \(\nabla_{{\bf u}} {\bf Q}\) is still a null matrix and therefore trivially of strict lower
triangular form. Therefore, this problem is of the class of equations
considered in \secref{sec:RedPrinc} and the reduction principle may be used in
order to define the appropriate resolvent variable and derive the resolvent
equation:
\begin{align}
\phi & \equiv \exp\left\{\int_0^t\!\displaystyle\left[\alpha\left(u_x^2+u_y^2\right)^k + \beta\right]{\rm d} \tau\right\} \\
u_x & = \frac{u_{x,0}}{\phi} \label{eq:gendragsolformx}\\
u_y & = \frac{1}{\phi}\left(u_{y,0}-g\displaystyle\int_0^t\!\phi\,{\rm d} \tau\right) \label{eq:gendragsolformy} \\
\frac{{\rm d} \phi}{{\rm d} t} & = \alpha\!\left[u_{x,0}^2 + \left(u_{y,0}-g\int_0^t\! \phi\,{\rm d} \tau\right)^2\right]^{k}\phi^{1-2k} + \beta \phi
\end{align}
Introducing \(\Phi\) according to \eqref{eq:order_reductio} yields an ordinary differential equation of second order:
\begin{equation}
\frac{{\rm d}^2\Phi}{{\rm d} t^2} = \alpha\left[u_{x,0}^2+\left(u_{y,0}-g\Phi\right)^2\right]^{k}\left(\frac{{\rm d} \Phi}{{\rm d} t}\right)^{1-2k} + \beta \frac{{\rm d} \Phi}{{\rm d} t}\;.
\end{equation}
The initial conditions remain the same as in the previous problem. A power series ansatz is now used, setting
\begin{equation}
\Phi = b_0 + b_1 t + b_2 t^2 + b_3 t^3 + \cdots
\end{equation}
The initial conditions require that \(b_0 = 0\) and \(b_1 = 1\). All further coefficients then follow from Taylor's theorem and may be computed recursively using
\begin{equation}
b_{n+2} = \frac{1}{\left(n+2\right)!}\left.\frac{{\rm d}^n}{{\rm d}
t^n}\left\{\alpha\left[u_{x,0}^2+\left(u_{y,0}-g\Phi\right)^2\right]^{k}\left(\frac{{\rm d} \Phi}{{\rm d} t}\right)^{1-2k} + \beta \frac{{\rm d} \Phi}{{\rm d} t}\right\}\right|_{t=0}.
\label{eq:gendragpowerseriessol}
\end{equation}
The first few terms of the solution hence read
{
\begin{align*}
b_0 & = 0 \\
b_1 & = 1 \\
b_2 & = \frac{\alpha u_0^{2k} + \beta}{2} \\
b_3 & = \frac{\left(1-2k\right)\alpha^2u_0^{4k} - 2k\,\alpha g u_0^{2k-1} \sin\theta + 2\left(1-k\right)\alpha\beta u_0^{2k} + \beta^2}{2\cdot 3} \\
\cdots & \quad\, \cdots \quad \cdots \quad \cdots \quad \cdots \quad \cdots \quad \cdots \quad \cdots \quad \cdots \quad \cdots \quad \cdots \quad \cdots
\end{align*}
}

\paragraph{Remark} The general nature of the coefficients is a bit more
transparent if one chooses units such that $u_0 = g = 1$. Then the coefficients
are bivariate polynomials in \(\alpha\) and \(\beta\). Each coefficient
comprises terms of three different kinds\added{.} First, there are terms \added{which}
depend only on \(\alpha\) and thus represent the effect of the general power
drag law, which is associated with high Reynolds numbers. Secondly, there are
terms \added{which} represent the effect of Stokes drag, i.e. low Reynolds numbers.
Other than these \emph{pure} terms, there are \emph{mixed} terms \added{which} are
responsible for the transition between these two regimes.
Although the underlying drag law is, formally, a linear superposition of a
general power law and a linear law, the solution itself is not a sum of the
solutions corresponding to the respective drag laws. As such, the existence of
the mixed terms illustrates the nonlinearity of the problem. Nonetheless, this
behaviour only becomes apparent from the third-order term onwards.

It is readily checked that upon setting \(\beta = 0\) and \(k = 1/2\), one
obtains the same expansion as established in \secref{sec:NewtonDrag} \added{above}. On the
other hand, for \(\alpha = 0\), one obtains \(\Phi = t + \beta t^2/2! + \beta^2
t^3/3! + \cdots = \left({\rm e}^{\beta t} - 1\right)/\beta\), whence \(u_x =
u_{x,0}{\rm e}^{-\beta t}\) and \(u_y = \left(u_{y,0} + g/\beta\right){\rm
e}^{-\beta t} - g/\beta\), which is the well-known closed-form solution for
Stokes drag (cf. Synge and Griffith \cite{Synge_Griffith1949}, p. 159). This serves as a
consistency check.

\subsection{Accounting for a Velocity Profile}
\label{sec:gen:subsec:prof}
\noindent As a final step of generalization, the effect of velocity
profiles on projectile motion \added{is considered}. After deriving the mathematical formulation,
i.e. the equations of motion, the solution is obtained using the result of
\secref{sec:RedPrinc}. Finally, some properties of the solution are pointed out.
\subsubsection{Equations of Motion}
\noindent The equations of motion for projectile motion subject to quadratic
drag \added{in a fluid of velocity \({\bf V}\)} are given by
\begin{equation}
\frac{{\rm d} {\bf v}}{{\rm d} t} = - \alpha \| {\bf v} - {\bf V} \| \left({\bf v} -
{\bf V}\right) + {\bf g}\;.
\end{equation}
The usual Cartesian coordinate system is chosen. Unlike in the previous
problems, however, the fluid velocity \added{\({\bf V}\)} is now allowed to vary with height. We
assume a unidirectional horizontal flow along the \(x\)\added{-}axis with the velocity
depending only on the vertical coordinate \(y\), i.e.
\begin{equation}
{\bf V} = \left(V_x(y),0\right)\;,
\end{equation}
so that \(V_x\) is a sufficiently well-behaved function of \(y\), subject to
the condition \(V_x(y=0) = 0\). Physically, this represents a flow in the
\(x\)-direction over a horizontal wall placed at \(y = 0\) with the fluid
obeying the no-slip condition of a viscous fluid at a solid wall. Here, we
restrict ourselves to the case where the velocity of the fluid varies linearly
with height, setting \(V_x = \gamma y.\) Such a law is valid for flows (both
laminar and turbulent) close to a smooth wall. In principle, a body moving in a
viscous shear flow experiences a torque proportional to its size due to the
velocity gradient. In the present context, the extension of the body is assumed
to be negligible compared to the velocity gradient, so that this effect can
safely be disregarded. The differential equations describing the motion may
then be written in coordinate form as
\begin{align}
\frac{{\rm d} v_x}{{\rm d} t} + \alpha\left(v_x - \gamma y\right)\sqrt{\left(v_x -
\gamma y\right)^2 + v_y^2} & = 0 \\
\frac{{\rm d} v_y}{{\rm d} t} + \alpha v_y \sqrt{\left(v_x - \gamma y\right)^2 + v_y^2} & = -g
\end{align}
The initial conditions are, as usual, \(v_x(0) = v_{x,0}\mbox{
}\textrm{and}\mbox{ }v_y(0) = v_{y,0}\). In addition, it is assumed that the
particle is released into the flow at \(x(0) = 0\) and \(y(0) = y_0 \geqslant
0\). Recalling that \(v_y = {\rm d} y/{\rm d} t\), one recognizes that due to the
occurence of the vertical position \(y\) the equations of motion actually
constitute, in their present form, a second-order system of coupled
differential equations (the second equation involving \({\rm d} v_y/{\rm d} t =
{\rm d}^2 y/{\rm d} t^2\)). The problem can, of course, be reformulated as a larger
system of first order (including \(y\) governed by \({\rm d} y/{\rm d} t = v_y\) as
the third component). Doing so, however, violates the structure exploited in
the reduction principle
\eqref{eq:abstractodeconstruction}--\eqref{eq:abstractodetoughsol} introduced
before in \secref{sec:RedPrinc}. This inconvenience can be resolved if the
problem is reformulated in terms of the relative velocity \({\bf u} = {\bf v} -
{\bf V} = \left(v_x - \gamma y, v_y\right) \equiv \left(u_x,u_y\right)\) yielding
\begin{align}
\frac{{\rm d} u_x}{{\rm d} t} + \alpha u_x\sqrt{u_x^2 + u_y^2} & = - \gamma u_y \label{eq:velprofuxcomp} \\
\frac{{\rm d} u_y}{{\rm d} t} + \alpha u_y\sqrt{u_x^2 + u_y^2} & = -g \label{eq:velprofuycomp}
\end{align}
subject to the initial conditions \(u_x(0) = u_{x,0}\) and \(u_y(0) =
u_{y,0}\).

\subsubsection{Solution}
\noindent In order to recast the system into the form considered in
\eqref{sec:RedPrinc}, the formulation of \(P\) may be adopted without change
from \eqref{eq:PQ_quadDrag}, since the same form of drag applies here as well.
On the other hand, as an effect of the \added{fluid} velocity profile, the inhomogenous term
now reads \({\bf Q} = \left(-\gamma v_y, -g\right)\). The gradient of \({\bf
Q}\) hence is
\begin{equation}
\nabla_{\!{\bf u}}{\bf Q} =
\left(\begin{array}{@{}cc@{}}
                    0 & -\gamma \\
                    0 & {\color{white} -}0 
\end{array}\right)\;,
\end{equation}
i.e. no longer zero. Therefore, the traditional methods as in \cite{Bernoulli1719} or
\cite{Parker1977} are inadequate. However, the problem is still within the scope of the
reduction principle from \secref{sec:RedPrinc}, because \(\nabla_{\!{\bf
u}}{\bf Q}\) is, up to numeration of indices, a strictly lower triangular
matrix. The procedure then yields
\begin{align}
\phi & \equiv \exp\left(\alpha\displaystyle\int_0^t\!\sqrt{u_x^2+u_y^2}\,{\rm d}\tau\right) \\
u_y & = \frac{1}{\phi}\left(u_{y,0} - g \displaystyle\int_0^t\! \phi \, {\rm d}\tau\right) \\
u_x & = \frac{1}{\phi}\left(u_{x,0} - \gamma \displaystyle\int_0^t \! u_y(\phi)\,\phi\, {\rm d}\tau \right) \\
& = \frac{1}{\phi}\left[u_{x,0} - \gamma \displaystyle\int_0^t\!\left(u_{y,0} - g \displaystyle\int_0^\tau \!\phi\,{\rm d}s\right){\rm d}\tau\right]
\end{align}
This may be further simplified by setting
\begin{equation}
\Psi \equiv \int \!\!\!\! \int \phi \, {\rm d}t\, {\rm d}t\;.
\label{eq:defpsi}
\end{equation}
The first initial condition is imposed by the definition of \(\phi = {\rm d}^2
\Psi/{\rm d}t^2\), namely
\begin{equation}
\frac{{\rm d}^2\Psi(0)}{{\rm d}t^2} = 1\;.
\end{equation}
Two other initial values are still necessary in order to determine \(\Psi\)
uniquely. This may be accomplished, e.g., by fixing the choice of integration
constants in \eqref{eq:defpsi}. The \added{need to satisfy the} physical initial
conditions \(u_x(0) = u_{x,0}\) and \(u_y(0) = u_{y,0}\) \added{suggests the choice}
\begin{align}
\Psi(0) & = \frac{u_{x,0}}{g\gamma} \\
\frac{{\rm d}\Psi(0)}{{\rm d}t} & = \frac{-u_{y,0}}{g}
\end{align}
Now, the solutions may be rewritten in terms of the auxiliary function \(\Psi\)
as
\begin{align}
u_x & = \,\gamma g\frac{\Psi}{{\rm d}^2\Psi/{\rm d}t^2} \\
u_y & = -g\frac{{\rm d}\Psi/{\rm d}t}{{\rm d}^2\Psi/{\rm d}t^2}
\end{align}
Inserting this into \eqref{eq:velprofuycomp}, yields
\begin{equation}
\frac{{\rm d}^3 \Psi}{{\rm d}t^3} = \alpha g \sqrt{\left(\gamma \Psi\right)^2 + 
\left({\rm d}\Psi/{\rm d}t\right)^2}\;. \label{eq:velprofred}
\end{equation}
This is the new auxiliary equation to which the original coupled system has
been reduced. One can now develop \(\Psi\) in a power series \(\Psi = c_0 + c_1
t + c_2 t^2 + c_3 t^3 + \cdots\) with
\begin{equation}
c_n = \frac{1}{n!}\frac{{\rm d}^n \Psi}{{\rm d}t^n}
\end{equation}
according to Taylor's theorem. The first three coefficients are given by the
initial conditions. Differentiation of \eqref{eq:velprofred} \(n\) times yields
the recursive formula
\begin{equation}
c_{n+3} = \frac{\alpha g}{(n+3)!}\left.\frac{{\rm d}^n}{{\rm
d}t^n}\sqrt{\left(\gamma \Psi\right)^2 + \left({\rm d}\Psi/{\rm
d}t\right)^2}\right|_{t=0},
\end{equation}
which can be used to evaluate all the coefficients of the series solution, the
first few terms reading
{\allowdisplaybreaks
\begin{align*}
c_0 & = \frac{u_0 \cos \theta_0}{g\gamma} \\
c_1 & = -\,\frac{u_0 \sin \theta_0}{g} \\
c_2 & = 1/2 \\
c_3 & = \frac{\alpha u_0}{2\cdot 3} \\
c_4 & = -\,\frac{\alpha \left[g \sin \theta_0 + \frac{1}{2}\gamma u_0
\sin(2\theta_0)\right]}{2\cdot 3\cdot 4} \\
\cdots & \quad\, \cdots \quad \cdots \quad \cdots \quad \cdots \quad \cdots \quad \cdots
\end{align*}}
The solution then is
\begin{align*}
u_x & = \frac{u_{x,0}}{1 + \alpha u_0 t + \cdots} - \gamma t \times \frac{u_0
\sin \theta_0 - \frac{1}{1\cdot 2} gt - \frac{1}{1\cdot 2\cdot 3} \alpha
u_0 g t^2 + \cdots}{1 + \alpha u_0 t + \cdots}
\\
u_y & = \frac{u_{y,0}}{1 + \alpha u_0 t + \cdots} - gt \times \frac{1 +
\frac{1}{1\cdot 2}\alpha u_0 t + \cdots}{1 + \alpha u_0 t + \cdots}
\end{align*}
Using the transformation equations given earlier, one can deduce \(v_x,v_y,x\)
and \(y\) from the above.

\subsubsection{Properties}

\paragraph{Comment on the  auxiliary variable} Here, in
\secref{sec:gen:subsec:prof}, the abstractly defined resolvent variable \(\Psi\)
no longer has the usual connotation of being (up to linear transformations) the
slope of the trajectory of the projectile. This also explains why the usual
approach (cf. \cite{Parker1977}) of assuming a resolvent variable linearly related to
\(u_y/u_x\) is not fruitful in this case. It may be of interest to note that
\begin{equation}
\frac{u_y}{u_x} = -\frac{1}{\gamma\,\Psi}\frac{{\rm d}\Psi}{{\rm d}t}\;,
\end{equation}
which is, up to constant factors, the logarithmic derivative of \(\Psi\). From
a similar point of view, the hodograph technique also appears to be problematic
because expressing \(\Psi\) in terms of $\tan \theta = u_y/u_x$ and inserting
the resultant expression into \eqref{eq:velprofred} would yield an
integro-differential equation involving non-trivial kernels.

\begin{figure}[h!]\sidecaption
\centering
\includegraphics[width = 0.5\textwidth]{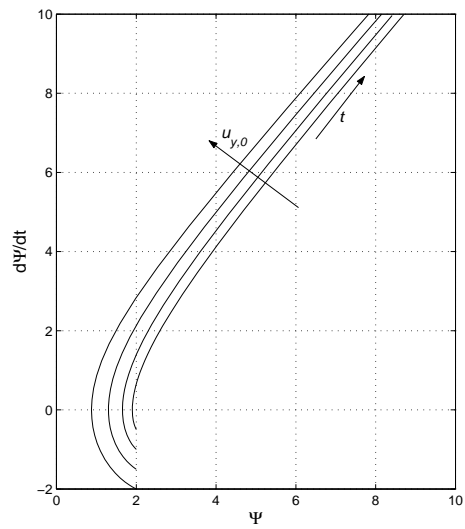}
\caption{Family of phase trajectories for $\Psi$ in normalized units such that $\alpha = g = 1$ and fixed $\gamma = 1$. The initial conditions are: fixed $u_{x,0} = 1$ and varying $u_{y,0} = 0.5,1,1.5,\mbox{ }\textrm{and}\mbox{ }2.$}
\label{fig:ProjMVelProfPhasePlot1}
\end{figure}

\paragraph{Particular Solutions} Although the complete solution could not be
expressed in closed form, surprisingly, there are several distinct non-trivial
particular solutions that may be expressed in terms of elementary functions. To
this end, assume an exponential ansatz of the form \(\Psi = C\exp\left(\lambda
t\right)\), where \(C\) is an arbitrary constant \added{and \(\lambda \in \mathbb{C}\)}. Inserting this into
\eqref{eq:velprofred} yields the characteristic equation
\begin{equation}
\lambda^3 = \alpha g \sqrt{\gamma^2 + \lambda^2}\;.\label{eq:chareq}
\end{equation}
for \(\lambda\). \added{Upon squaring this equation, one obtains a bi-cubic equation with solutions \(\lambda = \pm\lambda_i\) for \(i \in \left\{1,2,3\right\}\), whereby the values \(\mu_i = \lambda_i^2\) satisfy the cubic equation
\begin{equation}
\mu^3 = \alpha^2 g^2 \left(\gamma^2 + \mu\right)\;.
\end{equation}
Since the principle value of the square root in \eqref{eq:chareq} must be taken, this rules out one of the possible signs of \(\lambda = \pm\lambda_i\), so that \eqref{eq:chareq} indeed} has three roots in \(\mathbb{C}\)
corresponding to three separate solutions for the set of initial conditions
\(\Psi(0) = C, {\rm d}\Psi(0)/{\rm d}t = \lambda C\), \({\rm d}^2\Psi(0)/{\rm
d}t^2 = \lambda^2 C\). Since there is only one free parameter contained in the
solution, a general set of initial conditions cannot be fulfilled, neither can
the general solution be obtained by linear superposition owing to the
nonlinearity of the problem.
It would appear, nevertheless, that the properties of the particular solutions
reflect the behaviour of the general solution, so that knowledge of the former
would be instructive in comprehending the latter.
This may be visualized by means of phase plots in the \(\left(\Psi, {\rm
d}\Psi/{\rm d}t\right)\) plane. In \figref{fig:ProjMVelProfPhasePlot1}, a family
of phase trajectories is displayed for \(\gamma = 1\) and various initial
conditions, with \(\alpha = g = 1\) achieved by an appropriate choice of units.
The qualitative shape may indeed be derived from the nature of the particular
solutions found before. The
solutions of \eqref{eq:chareq} are such that one is positive real and two of
them are complex conjugates with negative real part. The complex conjugate
solutions are associated with exponentially damped sinusoidals, translating
into a spiralling ellipse in the phase plane. Their influence, however, decays
with time due to the damping. The real root, representing a growing exponential
function, quickly begins to dominate the solution, so that the long-term
behaviour of the solution in \figref{fig:ProjMVelProfPhasePlot1} is a straight
line. The slope is given by the real solution of \(\lambda^3 = \sqrt{\lambda^2
+ 1}\), i.e.
\begin{equation*}
\lambda = \frac{1}{\sqrt{3}}\sqrt{\sqrt[3]{\frac{27 - 3\sqrt{69}}{2}} + \sqrt[3]{\frac{27 + 3\sqrt{69}}{2}}} \approx 1.15096
\end{equation*}

\begin{figure}[h!]
\centering
\includegraphics[width = 0.65\textwidth]{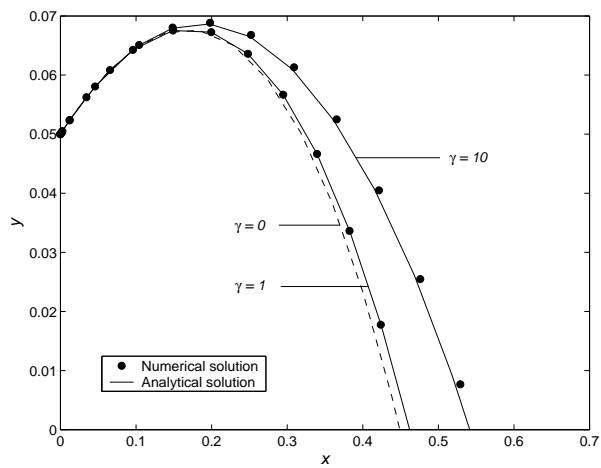}
\caption{Situation of a particle released at $y_0 = 0.05$ in units such that $\alpha = g = 1$ and with initial conditions $v_{x,0} = 1, v_{y,0} = 1/5$. Displayed are the trajectories for different values of the slope of the \added{fluid} velocity profile, $\gamma$. The solution obtained from truncating $\Psi$ after the fifth-order term (solid line) is compared with the numerical solution by a Runge-Kutta method. For reference, the Runge-Kutta solution without velocity profile ($\gamma = 0$) is also depicted (dotted line).}
\label{fig:ProjMVelProfExample}
\end{figure}
\clearpage
\section{Conclusions and Outlook}
In this paper, the motion of a body in a resistant medium and a constant
gravitational field was investigated. The nonlinear nature of the drag force
leads to a coupled nonlinear system of ordinary differential equations, thus
demanding a more involved approach in the analytical treatment. To this end, a
useful technique for reducing a particular class of coupled systems of ODEs
was proposed in order to facilitate the solution of such problems using
analytical means.

This \added{technique} was applied to projectile motion under quadratic drag, yielding an
equation of motion for a single auxiliary variable. On this basis, an exact
explicit solution to the problem using elementary operations on analytic expressions was
proposed. Relations to limiting cases from earlier studies were then shown. \added{Although an exact radius of convergence could not be provided so far, the good behaviour of the solution was demonstrated by considering partial sums with different truncations. In particular,} the
solution was also validated numerically by comparing it with direct solutions
of the original system of equations using a Runge-Kutta method. Furthermore, it was demonstrated that, when expressed in terms of
appropriate variables, the constant of motion for this problem \added{known in a different form from the literature} can be interpreted as the
conserved Hamiltonian of the system. Within this framework, it was further
pointed out that the resolvent variable provided by the reduction principle
together with its time derivative constitute the canonical or Darboux
coordinates of the motion.

The reduction technique was carried over to handle a more general drag law
containing quadratic and general power drag laws. A special case of this law is
the Schiller-Naumann correlation valid for the transition regime between
Stokesian and Newtonian drag. An exact explicit solution in the form of a ratio
of time-dependent power series was presented for this case as well.

Lastly, the reduction principle was also used in order to deal with another
nontrivial generalization of the basic problem, the motion of an object with
quadratic drag in shear flow with a linear velocity profile. This was motivated
by the fact that no study investigating projectile motion under the effect of a
\added{fluid} velocity profile and nonlinear drag could be found in the literature. Again, a
solution in the form of a ratio of power series was presented. Furthermore,
closed-form nontrivial particular solutions for the resolvent variable were
identified and the close relation between their properties and the nature of
the general solution was discussed. These could be the basis of new studies,
either from the perspective of the construction of a general solution from the
particular solutions (nonlinear superposition) or from the point of view of
semi-analytical approximations, e.g., using the method proposed by
Churchill and Usagi \cite{Churchill1972}.

While in practical applications a numerical solution of the class of systems of
ODEs considered here may be more efficient in computing a solution for given
initial conditions, the present results are important as they provide a means
of \added{investigating} the mathematical properties of the system and, ultimately, \added{gaining access to} more
physical insight, even in more general cases as demonstrated.

Several extensions of the present work are now possible:
The proposed reduction technique may be applied to other ODE systems
representing other phenomena, even possibly outside physics. An extension of
the method to other possibly more general classes of differential equations may
also be of interest.

Second, it seems appealing to generalize the shape of the background velocity
profile of the fluid to some classes of nonlinear functions. Finally, the
considered problem can be enhanced by considering other forces as well, of
which the Magnus force presumably is the most significant, requiring an
additional equation for the angular momentum of the particle.

\section*{Acknowledgements}
The first author would like to express his gratitude to
Martin-Andersen-Nex\"{o}-Gymnasium Dresden for providing an ideal framework
within the school curriculum for carrying out the research activities. The
authors are also thankful to Prof. Dr. Ralph Chill, Prof. Dr. J\"{u}rgen Voigt
and Prof. Dr. Herbert Balke for advice on mathematical background and
constructive criticism during the process of preparation of the manuscript as
well as Mr. Cornelius Demuth for proofreading the drafts.

\addcontentsline{toc}{section}{References}
\bibliography{manuscript2}

\begin{thebibliography}{27}
\providecommand{\natexlab}[1]{#1}
\providecommand{\url}[1]{\texttt{#1}}
\providecommand{\urlprefix}{}

\bibitem[{Arnol'd(1992)}]{Arnold1992}
Arnol'd, V.I.: Ordinary Differential Equations.
\newblock Springer Textbook Series, Springer Science \& Business Media, New
  York (1992), title of the original Russian: Obyknovennye differentsial'nye
  uravneniya, 3rd edition, Publisher Nauka, Moscow 1984; translated by Cooke,
  R.

\bibitem[{Benacka(2010)}]{Benacka2010}
Benacka, J.: Classroom notes: Solution to projectile motion with quadratic drag
  and graphing the trajectory in spreadsheets.
\newblock Int. J. Math. Educ. Sci. Technol. 41(3), 373--378 (2010)

\bibitem[{Benacka(2011)}]{Benacka2011}
Benacka, J.: On high-altitude projectile motion.
\newblock Can. J. Phys. 89(10), 1003--1008 (2011)

\bibitem[{Bernoulli(1719)}]{Bernoulli1719}
Bernoulli, J.: Responsio ad nonneminis provocationem, ejusque solutio
  quaestionis ipsi ab eodem propositae, de invenienda linea curva quam
  describit projectile in medio resistente.
\newblock Acta Eruditorum Maji, 216--226 (1719)

\bibitem[{de~Borda(1772)}]{Borda1769}
de~Borda, J.C.: Sur la courbe d\'{e}crite par les boulets \& les bombes, en
  \'{e}gard \'{a} la r\'{e}sistance de l'air.
\newblock In: Histoire de l'Acad\'{e}mie Royale des Sciences / Ann\'{e}e M.
  DCCLXIX / Avec les M\'{e}moires de Math\'{e}matique \& de Physique de la
  m\^{e}me Ann\'{e}e, Tir\'{e}s des Registres de cette Acad\'{e}mie, pp.
  116--121. L'Imprimerie Royale, Paris (1772)

\bibitem[{Churchill and Usagi(1972)}]{Churchill1972}
Churchill, S.W., Usagi, R.: A general expression for the correlation of rates
  of transfer and other phenomena.
\newblock AIChE 18, 1121--1128 (1972)

\bibitem[{Clift et~al.(1978)Clift, Grace, and Weber}]{Clift1978}
Clift, R., Grace, J.R., Weber, M.E.: Bubbles, Drops and Particles.
\newblock Academic, New York (1978)

\bibitem[{Didion(1860)}]{Didion1860}
Didion, I.: Trait\'{e} de balistique.
\newblock Mallet-Bachelier, Paris, 2nd edn. (1860)

\bibitem[{Dominici(2003)}]{Dominici2003}
Dominici, D.: Nested derivatives: a simple method for computing series
  expansions of inverse functions.
\newblock Int. J. Math. Math. Sci. 2003, 3699--3715 (2003)

\bibitem[{Euler(1745)}]{Euler1745}
Euler, L.: Neue Grunds\"{a}tze der Artillerie enthaltend die Bestimmung der
  Gewalt des Pulvers nebst einer Untersuchung \"{u}ber den Unterscheid des
  Wiederstands der Luft in schnellen und langsamen Bewegungen, aus dem
  Englischen des Hrn. Benjamin Robins \"{u}bersezt und mit den n\"{o}thigen
  Erl\"{a}uterungen und vielen Anmerkungen versehen von Leonhard Euler.
\newblock A. Haude, Berlin (1745)

\bibitem[{Hackborn(2006)}]{Hackborn2006}
Hackborn, W.W.: The science of ballistics: Mathematics serving the dark side.
\newblock Proc. Can. Soc. Hist. Phil. Math. 19, 109--119 (2006)

\bibitem[{Hardy(1940)}]{Hardy1940}
Hardy, G.H.: A Mathematician's Apology.
\newblock Cambridge University Press, Cambridge (1940), reprinted in 1992

\bibitem[{Hayen(2003)}]{Hayen2003}
Hayen, J.C.: Projectile motion in a resistant medium. {Part I}: exact solution
  and properties.
\newblock Int. J. Nonlin. Mech. 38(3), 357--369 (2003)

\bibitem[{Lagrange(1809)}]{Lagrange1809b}
Lagrange, J.L.: M\'{e}moire sur la th\'{e}orie g\'{e}n\'{e}rale de la variation
  des constantes arbitraires dans tous les probl\`{e}mes de la m\'{e}chanique.
\newblock M\'{e}moires de l'Institute national, classe des Sciences
  math\'{e}matiques et physiques 9, 257--302 (1809), reprinted in
  \emph{{\OE}uvres}, VI, pp. 771--804

\bibitem[{Lamb(1923)}]{Lamb1923}
Lamb, H.: Dynamics.
\newblock Cambridge University Press, London, 2nd edn. (1923)

\bibitem[{Lambert(1767)}]{Lambert1767}
Lambert, J.H.: M\'{e}moire sur la r\'{e}sistance des fluides avec la solution
  du probl\`{e}me ballistique.
\newblock In: Histoire de l'Acad\'{e}mie royale des sciences et belles-lettres
  \'{a} Berlin / {Ann\'{e}e} M. DCCLXV, pp. 102--188. Haude et Spener, Berlin
  (1767)

\bibitem[{Le~Gendre(1782)}]{Legendre1782}
Le~Gendre, A.M.: Dissertation sur la question de balistique propos\'{e}e par
  l'Acad\'{e}mie royale des sciences et belles-lettres de Prusse pour le prix
  de 1782.
\newblock G. J. Decker, Berlin (1782)

\bibitem[{Liao(2004)}]{Liao2004}
Liao, S.J.: Beyond Perturbation: Introduction to the Homotopy Analysis Method.
\newblock Chapman \& Hall, London (2004)

\bibitem[{Nalpanis et~al.(1993)Nalpanis, Hunt, and Barret}]{Nalpanis+1993}
Nalpanis, P., Hunt, J.C.R., Barret, C.F.: Saltating particles over flat beds.
\newblock Journal of Fluid Mechanics 251, 661--685 (1993)

\bibitem[{Newton(1687)}]{Newton1687}
Newton, I.: Philosophi{\ae} Naturalis Principia Mathematica.
\newblock S. Pepys, London, 1st edn. (1687)

\bibitem[{Parker(1977)}]{Parker1977}
Parker, G.W.: Projectile motion with air resistance quadratic in the speed.
\newblock Am. J. Phys. 45(7), 606--610 (1977)

\bibitem[{Schiller and Naumann(1933)}]{SchillerNaumann1933}
Schiller, L., Naumann, A.: A drag coefficient correlation.
\newblock VDI Zeitschrift 77, 318--320 (1933)

\bibitem[{Synge and Griffith(1949)}]{Synge_Griffith1949}
Synge, J.L., Griffith, B.A.: Principles of Mechanics.
\newblock McGraw-Hill Book Company, Toronto, 2nd edn. (1949)

\bibitem[{Tsuboi(1996)}]{Tsuboi1996}
Tsuboi, K.: On the optimum angle of takeoff in long jump.
\newblock Trans. Jap. Soc. Ind. Appl. Math 6, 393 (1996)

\bibitem[{Walter(2000)}]{Walter2000}
Walter, W.: Gew\"{o}hnliche Differentialgleichungen / Eine Einf\"{u}hrung.
\newblock Springer-Verlag, Heidelberg, 7th edn. (2000)

\bibitem[{Weinacht et~al.(2005)Weinacht, Cooper, and Newill}]{Weinacht+2005}
Weinacht, P., Cooper, G.R., Newill, J.F.: Analytical prediction of trajectories
  for high-velocity direct-fire munitions.
\newblock Tech. Rep. ARL-TR-3567, U.S. Army Research Laboratory (2005)

\bibitem[{Yabushita et~al.(2007)Yabushita, Yamashita, and
  Tsuboi}]{Yabushita2007}
Yabushita, K., Yamashita, M., Tsuboi, K.: An analytic solution of projectile
  motion with the quadratic resistance law using the homotopy analysis method.
\newblock J. Phys. {A}-Math. Theor. 40, 8403--8416 (2007)

\end{thebibliography}
\bibliographystyle{splncsnat}

\end{document}